\documentclass[newversion]{aa}
\usepackage{graphicx}

\begin{document}


   \title{The Spatial Structure of the Evershed Effect}

   \author{E. Wiehr}

   \offprints{E. Wiehr}

   \mail{ewiehr@astrophysik.uni-goettingen.de}

   \institute{Institut f\" ur Astrophysik der Universit\"at,
              Friedrich-Hund-Platz 1, 37077 G\"ottingen, Germany}

   \date{Received Feb. 19, 2008; accepted July 9, 2008}

\abstract
{}
{The spatial disappearance of the line asymmetry from the Evershed effect 
near the boundary of the white-light penumbra is investigated.}
{The neighboring lines Ni I 5435.9 (g=0.5) and Fe I 5434.5 (g=0; formed 
about 300 km higher) are observed in a sunspot penumbra at $\vartheta = 65^o$.}
{Immediately beyond the ends of the dark penumbral continuum structures, both 
lines simultaneously lose their asymmetry. The spatial distance of their
disappearance of maximally 500 km is too short for a 'disappearance with 
height' as is suggested by models of a penumbral 'canopy'. Instead, the 
data favor a rather flat orientation of the Evershed flow with an abrupt 
disappearance. It is suggested that this location marks the sharp threshold 
of the equipartition between kinetic and magnetic energy density at the 
outer penumbral border.} 
{}

\maketitle

%
%

\section{Introduction}

The characteristic signature of the Evershed effect is a line profile 
asymmetry which often occurs as a pronounced line 'kink' (Maltby 1964) 
and can not be explained by a velocity gradient with depth (Stellmacher 
\& Wiehr 1980). Instead, one has to consider that the observed line profiles 
represent superpositions of spatially unresolved penumbral structures: a 
stronger 'main component' originating from the bright structures superposed 
by a Doppler-shifted and weaker line 'satellite' from the dark structures 
(Stellmacher \& Wiehr 1971). This is supported by the finding that pronounced 
line 'kinks' only occur if the (dark) penumbral structures are oriented almost 
along the line-of-sight, i.e. in center-side penumbrae of spots very close 
to the limb (Wiehr 1995). 

Observed differences between the Evershed effect 
in center- and limb-side penumbrae suggest angles of inclination to the solar 
surface between $6^o$ (Maltby 1964) and $15^o$ (Schr\"oter 1965). Such flat 
structures cannot gradually 'disappear with height' as is to be expected for 
a penumbral 'canopy' (cf. Solanki et al. 1992). In that picture the profile 
asymmetry of Fe\,I\,5434\AA{} (formed about 500 km above the continuum level), 
should 'disappear' over a horizontal distance of several arcsec beyond the 
white-light penumbral border. This, however, disagrees with the results of 
Wiehr \& Degenhardt (1992,1994). 

On the other hand, the chromospheric H$\alpha$ superpenumbra extends far 
beyond the edge of the continuum penumbra. But the opposite sign of the 
chromospheric Evershed effect (e.g. Maltby 1975) indicates that it cannot 
represent a continuation of the photospheric Evershed effect at higher 
levels. Hence, a possible extension of structures, flow, and magnetic 
field beyond the white-light penumbra needs further study. The present 
paper investigates an upper limit for the abrupt disappearance and its 
physical reason. 

\section{Observations}

On October 22, 1994, the neighboring lines Fe\,I\,5434.5 (g=0) and 
Ni\,I\,5435.9 (g=0.5) were observed with the Gregory telescope on Tenerife 
(Wiehr 1986) in a sunspot at a heliocentric angle of $\vartheta = 65^o$
with a 1024 x 1024 pixel CCD at an exposure time of 0.1\,seconds. In order 
to avoid the use of an image rotator, the time of observation was chosen 
such that the Coud\'e rotation of the solar image yields an almost parallel 
orientation of the slit to the limb nearest the spot. During the two hours 
of observation the image rotation increased the angle between slit and limb 
from $-15^o$ to $+15^o$ with respect to the exactly parallel orientation. 

An accurate pointing of the 40\,$\mu$m slit at very small distances from 
the outer penumbral edge (e.g. $\Delta<0.5$ arcsec) is difficult, since the 
slit occurs as a 1/3 arcsec wide dark streak on the slit-jaw image just at 
the location under study. However,  if the telescope's guiding automat
is switched off, a slow image drift of about 0.5 arcsec/min occurs, which 
affects a kind of automatic scanning if a large number of CCD-spectra is 
taken at a short time cadence. This yields spectra at a variety of slit 
distances including such of few 100\,km inside or outside the penumbral 
edge. Since the image rotation slowly varies the slit orientation on the 
sun, a repetition of such a series of 'drifted spectra' gives slit 
orientations of different azimuth angles with the spot border, thus 
covering large parts of the limb- and of the center-side penumbral 
boundaries. In the best spectra, a spatial resolution near 0.5\,arcsec 
is achieved. Two examples are shown in figure\,1.

   \begin{figure}[ht]     
   \hspace{-4mm}\includegraphics[width=9.9cm]{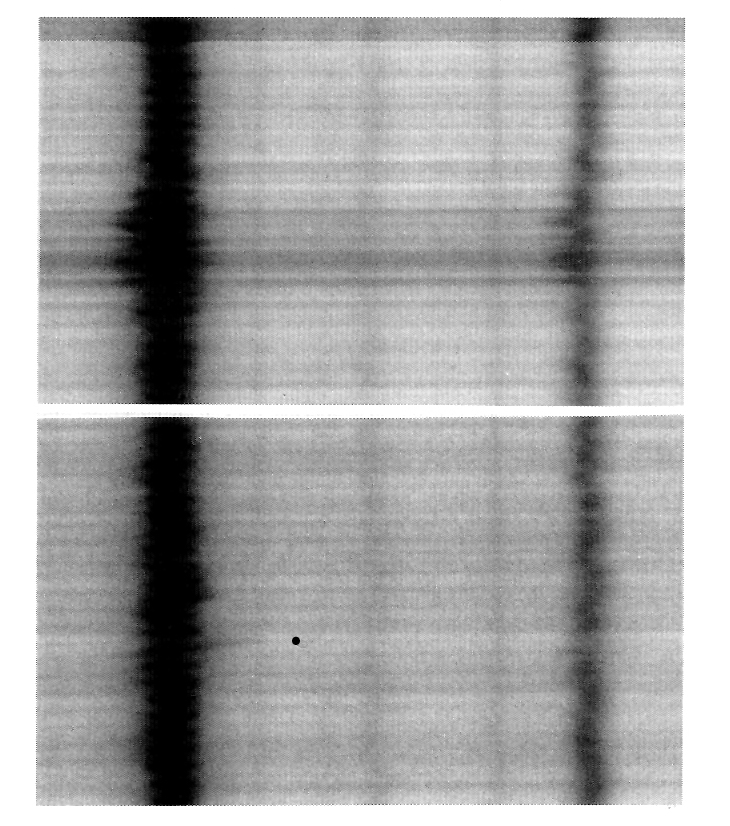}
   \caption{CCD spectra of Fe\,I\,5434.5 (left) and Ni\,I\,5435.9 (right) 
a) immediately inside the center-side, (b) just outside the limb-side 
penumbral border; the spatial extension of each spectrum covers 50\,arcsec.
The black dot marks a profile asymmetry {\it not} related to the Evershed 
effect.}
   \label{Fig1}
    \end{figure}

%
%
                      
%

   \begin{figure}[htb]     
   \hspace{0mm}\includegraphics[width=9.2cm]{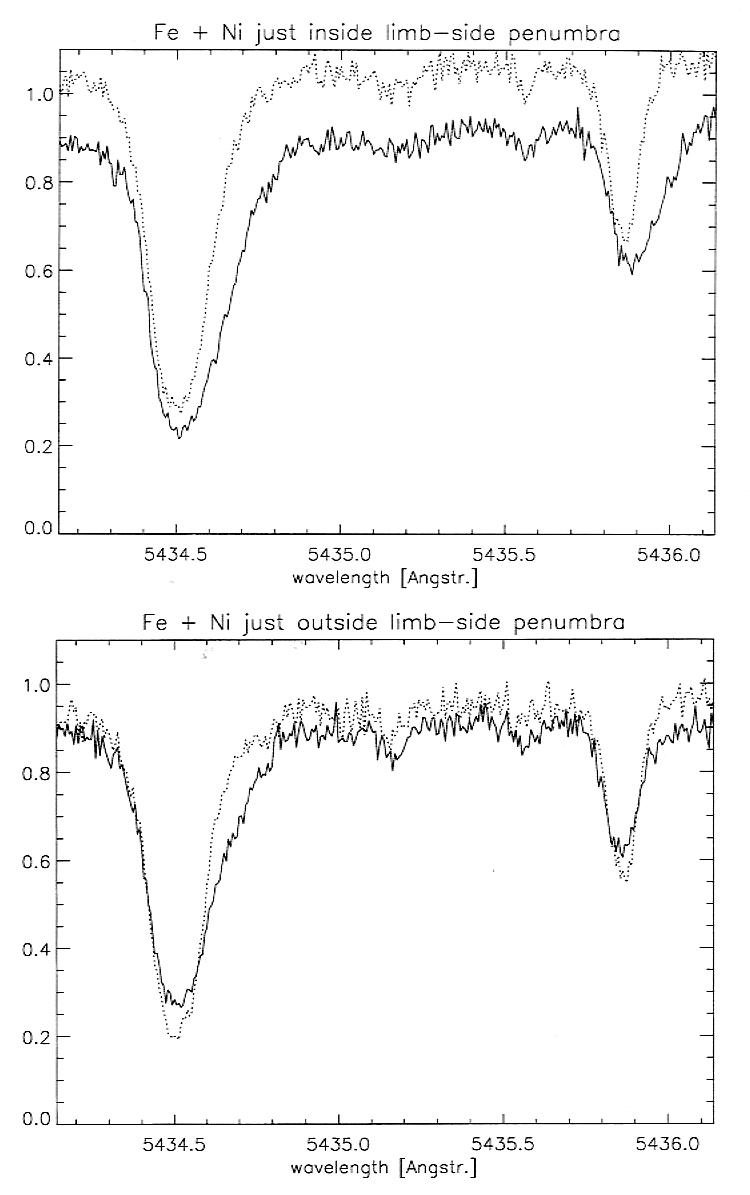}
   \caption{Fe\,I\,5434.5 and Ni\,I\,5435.9 with profile asymmetry 
at a location of a continuum depression produced by the end of a dark 
penumbral structure (solid line in the upper panel) in comparison with a 
location immediately outside the penumbra (dotted line). Lower panel: 
profile asymmetry only for Fe\,I\,5434.5 and not related with a continuum 
depression (solid line), originating from a high velocity region (dot in 
Fig.\,1b) outside the spot.}
   \label{Fig2}
    \end{figure}

%
%

\section{Results}

The numerous spectra obtained from slit positions in the immediate 
vicinity of the penumbral border confirm (e.g. Wiehr \& Degenhardt 1994) 
that line asymmetries occur exclusively at locations of pronounced 
continuum depressions produced by dark penumbral structures (solid line 
in Fig.\,2a). In the neighboring bright structures, both line profiles 
are symmetric (dotted lines in Fig.\,2). {\it Immediately outside the 
penumbra, where the slit just 'missed' the outer ends of the dark 
structures, the line profile asymmetries disappear simultaneously for 
both lines.} 

The spatial distance between adjacent positions of the slit is partly
determined by slow image drift and mainly by the image motion. The latter 
can be expected to be slightly smaller than the 0.5\,arcsec spatial 
resolution in the spectra, which are additionally degraded by the 
spectrograph optics. Assuming a spatial distance of the order of the 
slit width of 1/3\,arcsec and correcting this value for the geometric 
foreshortening at $\vartheta=65^o$, one obtains a {\it horizontal distance 
of maximally 500\,km over which the line profile asymmetries disappear 
simultaneously for the two lines, which formed at a vertical distance
of about 300\,km}.  

Among the numerous spectra showing a strictly simultaneous disappearance 
of the asymmetry of both lines, only one location is found (dot in Fig.\,1b) 
which exhibits a line asymmetry of Fe\,I\,5434 but not of Ni\,I\,5436 
(see Fig.\,2b). However, this structure shows no continuum depression 
and is thus not a dark penumbral filament. It is located between a bright 
continuum streak from a facula and another streak with opposite line 
asymmetry (cf. Fig.\,1b) thus indicating a disturbed region at the penumbral 
border (possibly an Ellerman bomb).

\section{Discussion}

The simultaneous disappearance of the profile asymmetry of both lines 
together with the continuum depressions supports the idea that the profile 
asymmetries are due to superposed line satellites from spatially unresolved 
dark penumbral structures (Wiehr 1995). These line 'satellites' disappear 
in photospheric layers together with the dark structures where they are 
formed. If a 'canopy' existed, one would expect the asymmetry of Fe\,I\,5434, 
(a line formed near the temperature minimum; see e.g. Degenhardt \& Wiehr 
1994)) to disappear further away from the penumbral continuum border than 
that of Ni\,I\,5436 (formed about 300\,km higher). In the 'canopy' scenario,
the simultaneous disappearance over a horizontal distance of less than 500\,km 
would require a steep inclination of the flow (and thus of the magnetic 
field) of at least $60^o$, which severely conflicts with the values observed 
by Maltby (1964) and by Schr\"oter (1965).

The present observations, however, strongly suggest a nearly horizontal 
orientation of the Evershed flow (and thus also of the magnetic field in the
dark penumbral structures) with an abrupt disappearance at the very border
of the white-light penumbra. A plausible explanation for such a sharp end is 
the equipartition between the magnetic and kinetic energy densities: For a 
photospheric density value of $\rho = 5\cdot 10^{-7}$ g/cm$^2$, a granular 
velocity of 3\,km/s balances 740\,Gs, which is close to values observed at 
the penumbral edge. If the magnetic field varies radially with $B\sim r^{-2}$ 
(cf. Wittmann 1974), the magnetic energy density $B^2/8\pi$ decreases with 
$r^{-4}$. This results in a sharp threshold for an equipartition with the 
kinetic energy density of the neighboring granules. Flow and field will 
than (vertically) sink at the very penumbral border.[Such a 'disappearance' 
of the flow may conflict with the siphon model (Meyer \& Schmidt 1968) but 
hardly with the idea of 'turbulent pumping' (Parker 1974).]

A 'canopy' formed by such an extension of the bright penumbral structures 
does not contradict R\"uedi et a1. (1995) who deduced significantly 
steeper inclinations from the He\,I\,line at 10830\AA{} than expected 
from a neighboring Si\,I\,line. The latter, however, is formed in deep 
layers which are dominated by flat dark structures. The helium line, 
instead, originates from high layers, which are dominated by the steep 
bright structures.

%
%

\begin{acknowledgements}
The author acknowledges numerous discussions with C.\,R. 
de\,Boer, G.\,Stellmacher, H.\,Balthasar, M.\,Sch\"ussler and a critical 
review of the manuscript by C.\,Zwaan.    
\end{acknowledgements}

%
%

\eject

%
%

\end{document}